\documentclass[twocolumn]{aastex631}
\usepackage{epsfig}
\usepackage{rotating}
\usepackage[T1]{fontenc}
\usepackage{ae,aecompl}
\usepackage{tikz}

\usepackage{amsmath}
\usepackage{amssymb}
\usepackage{graphicx}
\usepackage{color}
\usepackage{natbib}

\usepackage{commath}
\usepackage{hyperref}

\newcommand{\Ms}{{\ensuremath{M_{\odot} }}}
\newcommand{\Zs}{{\ensuremath{Z_{\odot} }}}

\submitjournal{AAS journals}

\shorttitle{DCBH growth}
\shortauthors{Latif et al.}

\begin{document}

\title{How Overmassive Black Holes Formed at Cosmic Dawn}

\correspondingauthor{Muhammad A. Latif}
\email{latifne@gmail.com}

\author[0000-0003-2480-0988]{Muhammad A. Latif}
\affiliation{Physics Department, College of Science, United Arab Emirates University, PO Box 15551, Al-Ain, UAE}

\author[0000-0001-6646-2337]{Daniel J. Whalen}

\affiliation{Institute of Cosmology and Gravitation, Portsmouth University, Dennis Sciama Building, Portsmouth PO1 3FX}

\author{Sadegh Khochfar}
\affiliation{Institute for Astronomy, University of Edinburgh, Royal Observatory, Blackford Hill, Edinburgh EH9 3HJ, UK}

\author{Fergus Cullen}
\affiliation{Institute for Astronomy, University of Edinburgh, Royal Observatory, Blackford Hill, Edinburgh EH9 3HJ, UK}

\begin{abstract}

Overmassive black hole galaxies (OBGs) at redshifts $z \sim$ 10, or 450 Myr after the Big Bang, are one of the most puzzling discoveries by the James Webb Space Telescope to date because they formed by such early epochs and their black-hole to stellar mass ratios are a hundred times higher than those in galaxies today.  Here we show that OBGs are simply the result of DCBH birth in primordial halos at early times.  A 70,000 \Ms\ DCBH forming at $z =$ 25.7 in our cosmological simulation grows at about half the Eddington rate to $6.0 \times 10^6$ \Ms\ by $z =$ 10.1.  Its host galaxy reaches a stellar mass of $4 \times 10^8$ \Ms, a metallicity $Z =$ 0.1 \Zs, a star formation rate of 2 \Ms\ yr$^{-1}$, and $M_{\rm BH}/M_{\ast}$  $\sim$ 0.01, on par with OBGs like GN-z11, UHZ1, and GHZ9 at $z =$ 10.6, 10.1, and 10.2, respectively.  Our simulation, the first to follow the coevolution of a DCBH and its host galaxy for several hundred Myr while resolving star formation in the earliest minihalos, shows that this ratio is a natural result of initial suppression of star formation by the DCBH and the later, violent blowout of metals by Pop III supernovae.  Our models yield an excellent match to the spectra of UHZ1 and GHZ9 at $z =$ 10.1 and 10.4, respectively.

\end{abstract}


\section{Introduction}

% Fig. 1

\begin{figure*}
\begin{center}
\includegraphics[scale=0.5]{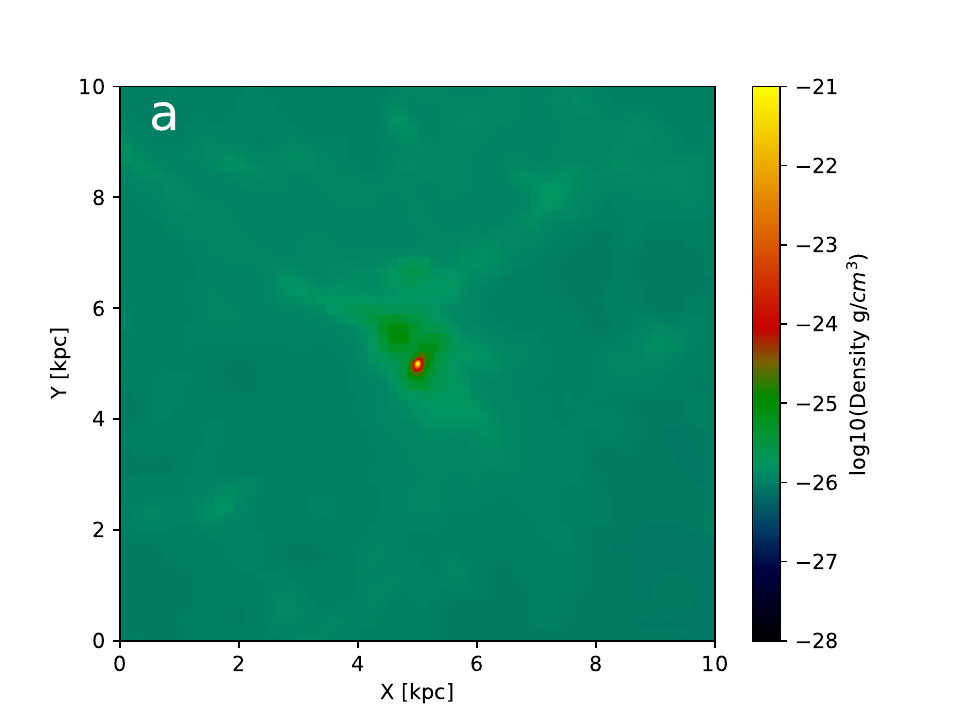} 
\includegraphics[scale=0.5]{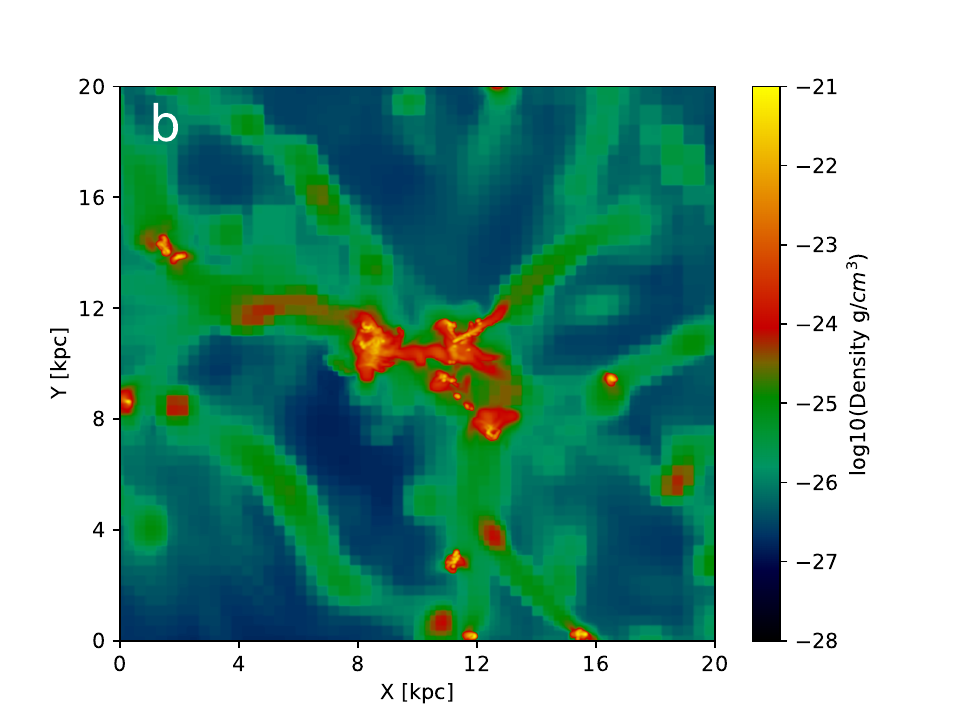} 
\includegraphics[scale=0.5]{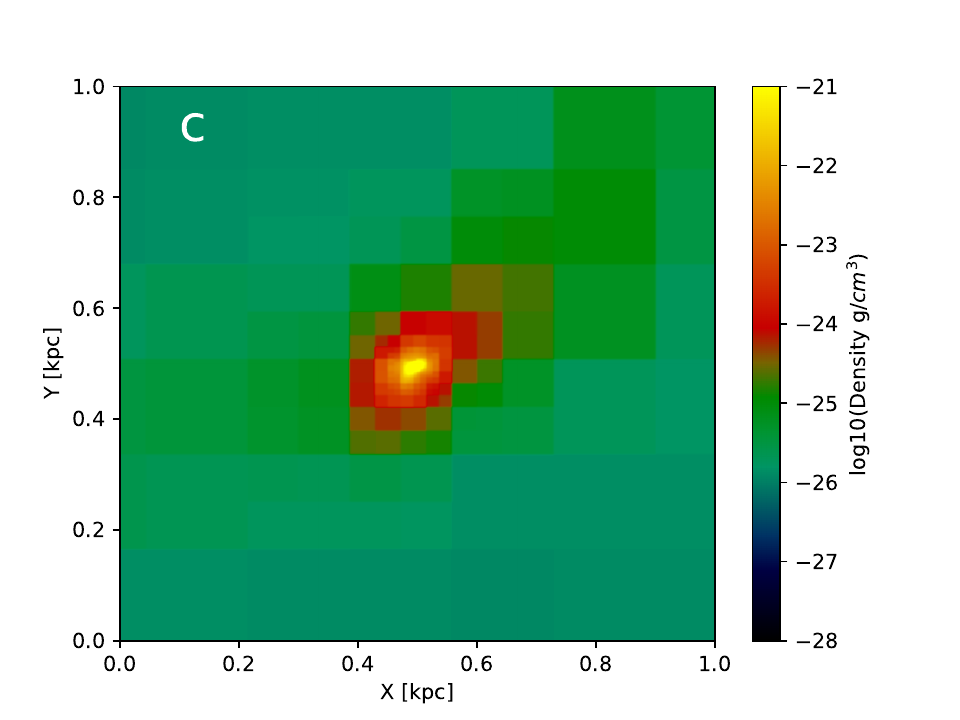} 
\includegraphics[scale=0.5]{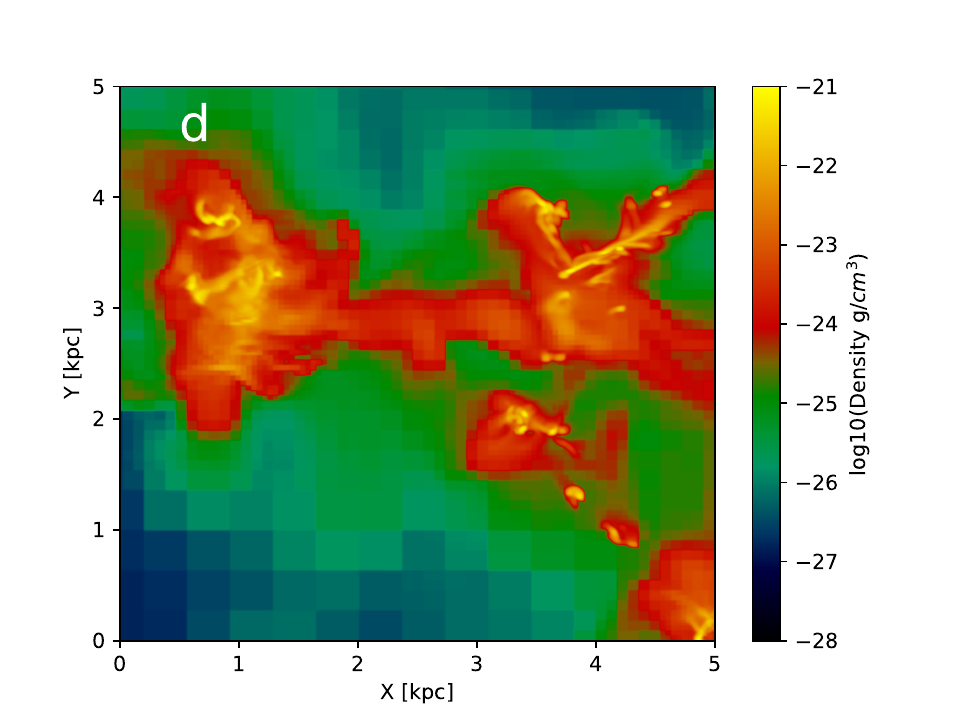} 
\end{center}
\caption{Panel (a): 10 kpc image of the DCBH at birth at $z =$ 25.7. Panel (b): 20 kpc image of the OBG at $z = 11.4$ centered on the black hole.  Panel (c):  1 kpc image of the DCBH at $z =$ 25.7.  Panel (d):  5 kpc image of the OBG at $z = 11.4$.}
\label{fig:obg}
\end{figure*}

Overmassive black hole galaxies (OBGs) at $z \gtrsim$ 10 are the new frontier of high-redshift supermassive black hole (SMBH) astronomy.  UHZ1, a $4 \times 10^7$ \Ms\ BH at $z = 10.1$ \citep{cet23,Bod23,Gould23}, GNz11, a $10^6$ \Ms\ BH at $z = 10.6$ \citep{B23,Maio24N} and GHZ9, an $8\times10^7$ \Ms\ BH at $z =$ 10.4 \citep{Atek23,Cast23,Kov24} are the first AGNs to be observed above $z =$10.  These BHs reside in relatively metal-poor galaxies and are about two orders of magnitude overabundant relative to the quasar population at similar redshifts \citep{G24}.  The host galaxies of UHZ1 and GHZ9 have stellar masses of $\sim 10^8$ \Ms\ so the BHs lie $\sim$ 2 orders of magnitude above the local $\rm M_{\rm BH} - M_{\star}$ relation \citep{rv15}. 

OBGs impose the strongest constraints on the seeds of high-$z$ SMBHs to date. Numerical simulations of Pop III star BH seeds of $\sim$ 100 \Ms\ with radiation transport show that it is nearly impossible for them to sustain such growth \citep{Smith18,latif20b,Gord25}.  Massive seeds of $10^4- 10^5$ \Ms\ due to the collapse of supermassive primordial stars \citep{hos13,tyr17,hle18b,nan25a} created by atomically-cooled halos \citep{rh09b,latif13c,bec18,pat23} or stellar collisions \citep{Yaji16,Sch22,Rein25,chon26} can reach observed BH masses by this epoch at accretion rates below the Eddington limit \citep{Agar13,Bod23,Kov24} but these estimates are based on analytical models.  Simulations of DCBH birth and growth in cosmological environments have reproduced some of the properties of OBGs \citep{jeon25,bhow26} but did not resolve star formation in minihalos, which is important for feedback onto the BH on small scales, and treated luminosity from the BH as heat rather than X-rays, which can catalyze Pop III star formation on small scales \citep[e.g.,][]{mba03}.  Cosmological simulations in other contexts with X-rays from the BH and ionizing UV, winds and supernovae due to Pop III and Pop II stars find that X-ray feedback stifles BH growth and that duty cycles average $\sim$ 50\% \citep{j11,aycin14,L18,smidt18,latif20b} but could not resolve these processes in the vicinity of the BH deep in the halo.

Here we investigate the formation of an OBG at cosmic Dawn with the first cosmological simulation to follow the coevolution of a DCBH and its host galaxy for several hundred Myr with radiative feedback from the BH and chemical, mechanical, and radiative feedback from Pop III and Pop II stars.  Our dark matter (DM) resolution of 3600 \Ms\ allows us to resolve star formation in minihalos of the host galaxy at $z \gtrsim$ 15.  We describe our numerical methods and simulation setup in Section 2 and discuss the evolution of the OBG in Section 3.  We calculate spectra for the OBG at $z =$ 10 and compare them to those of GHZ9, UHZ1 and GNz-11 in Section 4 and conclude in Section 5.

\section{Numerical Methods} \label{sec:method}

% Fig. 2

\begin{figure}
\begin{center}
\begin{tabular}{c}
\includegraphics[scale=0.4]{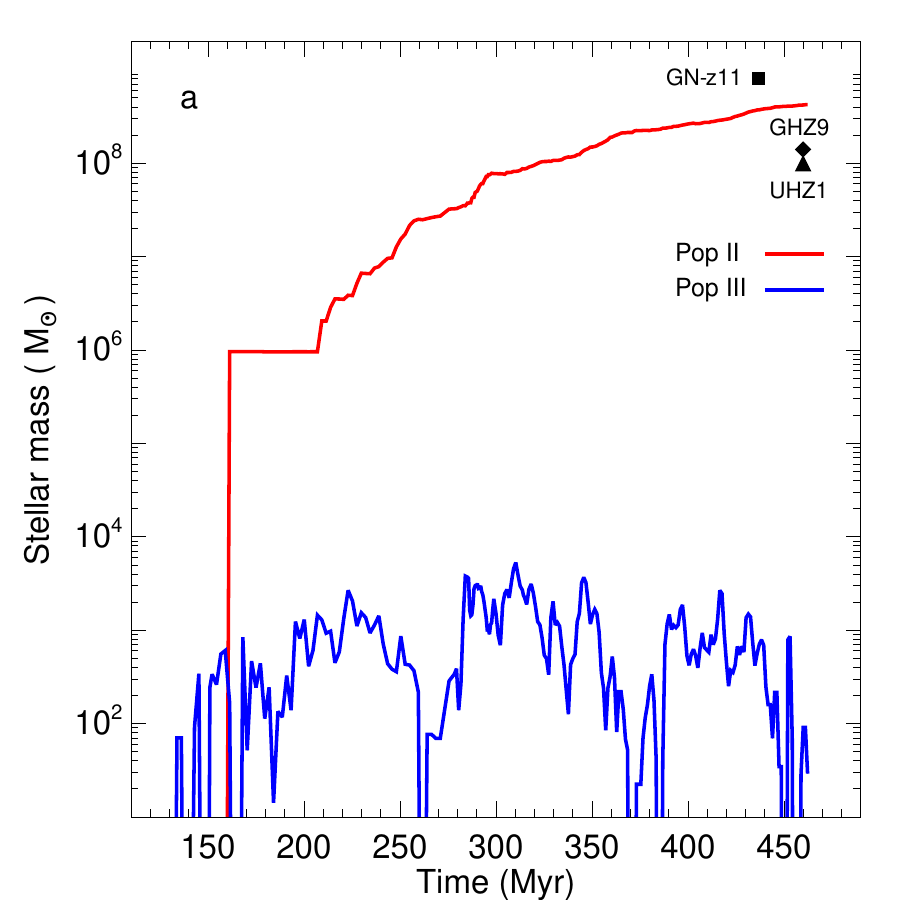}   \\
\includegraphics[scale=0.4]{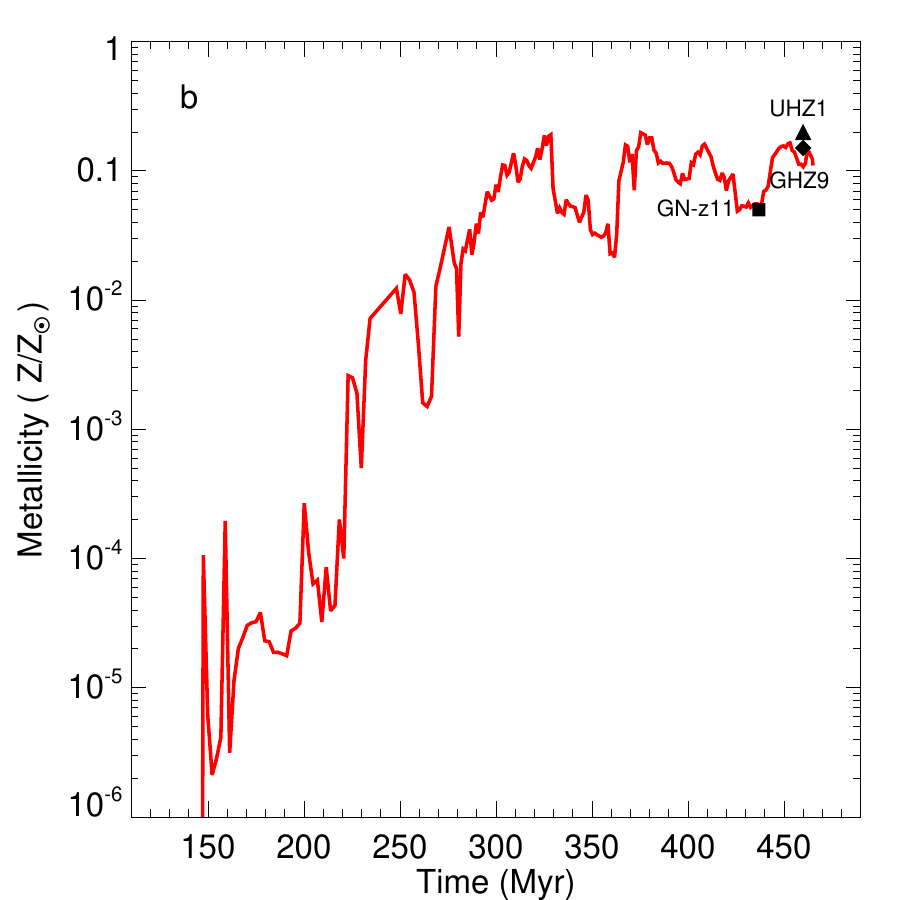}   \\
\includegraphics[scale=0.4]{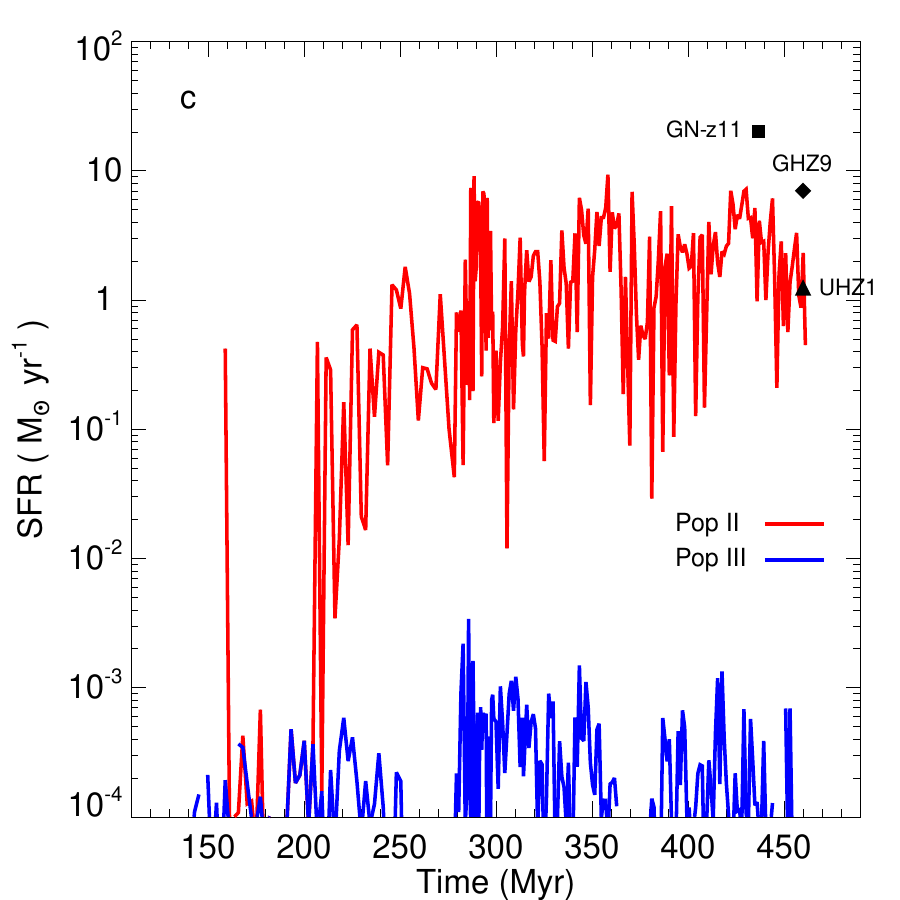} 
\end{tabular}
\end{center}
\caption{Stellar masses (a), average metallicity (b) and SFRs (c) in the host galaxy of the DCBH over cosmic time in Myr after the Big Bang.}
\label{fig:SF}
\end{figure}

We simulated the growth of the DCBH in \textsc{ENZO}, an adaptive mesh refinement (AMR) cosmology code that evolves gas, dark matter, ionizing UV and X-rays, and nonequilibrium primordial gas chemistry and cooling to model structure formation in the early Universe \citep{enzo}.  It uses a third-order piecewise-parabolic method for gas dynamics \citep{bryan95} with an HLLC Riemann solver for enhanced stability with strong shocks and rarefaction waves \citep{toro94}.  Dark matter dynamics is evolved with an $N$-body particle-mesh method \citep{efs85,couch91} and self-gravity is computed with a multigrid Poisson solver.  Nonequilibrium primordial gas reactions are self-consistently coupled to hydrodynamics by isochoric updates to the gas energy equation due to chemical cooling and heating.  The reaction network \citep{anet97} evolves mass fractions for $\rm H$, $\rm H^+$, $\rm He$, $\rm He^+$, $\rm He^{++}$, $\rm H_2$, $\rm H^-$, $\rm H_2^+$, and $\rm e^-$, and we include cooling due to collisional excitation and ionization of H and He, recombinations in H and He, bremsstrahlung emission, $\rm H_2$ ro-vibrational transitions, and inverse Compton scattering.  The H$_2$ cooling rates \citep{ga08} are valid at low densities and high densities, at which local thermodynamic equilibrium (LTE) effects become important.  They also include the transition from optically-thin to optically-thick cooling at densities when H$_2$ becomes opaque to its own cooling lines \citep{ra04}, although in practice we never reach this regime in our simulation.  Metal-line cooling \citep{japp07} from C, O, N, and Si is calculated from metallicity-dependent rates for $T = 100 - 10^4$ K and tabulated cooling functions \citep{Sutherland93} at $T \gtrsim 10^4$ K.  

Star particles are created in cells that meet the following criteria \citep{latif20b}:  (1) an overdensity of $5 \times 10^5$ (corresponding to $\rm 10^3~cm^{-3}$ at $z = 10$);  (2) an $\rm H_2$ mass fraction $\geq 5 \times 10^{-4}$; (3) a convergent flow ($\nabla \cdot v_{\rm gas} < 0$).  Pop III stars form when $\rm Z/Z_{\odot} \leq 10^{-4}$ and Pop II stars form otherwise, without requiring molecular hydrogen \citep{schn06,osh08}.  The formation of Pop III star particles is restricted to cells with $T <$ 1000 K.  Each Pop III star particle represents a single star whose mass is randomly drawn from an initial mass function (IMF) from 1 - 300 \Ms\ that follows a Salpeter slope above 100 \Ms\ and an exponential cutoff below that mass \citep{Wise12}.  Each Pop II star particle represents a small 10$^6$ \Ms\ stellar cluster.  Runs with less massive 10$^3$ \Ms\ star particles produced similar results. 

We propagate X-ray and ionizing UV photons through the simulation volume with the MORAY \citep{moray} ray tracing radiation transport code.  Heating and radiation pressure due to photoionizations are included in updates to the gas energy and momentum equations, respectively.  Energy from electrons due to X-ray photoionizations is partitioned between thermalization in the gas and secondary ionizations \citep{Shull85,kim11}.  Pop III and Pop II star particles are treated as point sources of ionizing photons with energies of 29.6 eV and 21.6 eV, respectively.  We use mass-dependent luminosities for Pop III stars \citep{s02} but take Pop II star particles to emit 6000 photons per stellar baryon over 20 Myr (equivalently, $\rm 2.4 \times 10^{47}~photons~s^{-1}$\Ms$^{-1}$).  Pop III SN feedback includes core-collapse (CC) and pair-instability (PI) explosions, depending on stellar mass \citep{hw02,het03}, and is modeled by injecting the kinetic energy of the blast as a free expansion into a 10 pc sphere. Pop II SN feedback is modeled by injecting $\rm 6.8 \times 10^{48}~erg~s^{-1}$\Ms$^{-1}$ within a 10 pc sphere over a period of 4 Myr after SF.

The BH is treated as a sink particle and a point source of X-ray photons with energies of 2 keV, consistent with the average quasar spectral energy distribution \citep{sos04} and observations that $\sim$ 90\% of the X-ray flux from quasars is 0.5 - 2 keV \citep{nan17}.  We assume Bondi-Hoyle accretion \citep{latif20b}:
\begin{equation}
\dot{M}_{\rm BH}  =  min \left( \frac{4 \pi G^2  M_{\rm BH}^2 \rho_{\rm B}}{c_{\rm s}^3},  \frac{4 \pi G M_{\rm BH} m_{\rm p}}{\epsilon_{\rm r} \sigma_{\rm T} c} \right)
\end{equation}
where $\dot{M}_{\rm BH}$ is the accretion rate, $\epsilon_{\rm r} = 0.1$ is the radiative efficiency, $G$ is the gravitational constant, $M_{\rm BH}$ the black hole mass, $\rho_{\rm B}$ is the density at the Bondi radius, $c_{\rm s}$ is the sound speed, $m_{\rm p}$ is the proton mass and $\sigma_{\rm T}$ is the Thomson scattering cross-section. The Bondi accretion radius is
\begin{equation}
R_{\rm B} = \frac{2 G M_{\rm BH}}{c_{\rm s}^2}.
\label{eq3}
\end{equation}
Our minimum cell size of 4 pc resolves the Bondi radius of a $10^5~ \Ms$ BH. The luminosity of the BH is then $L_{\rm{MBH}}= \epsilon_{\rm r} \dot{M}_{\rm{BH}} c^2$, where $c$ is the speed of light.     

The initial conditions of this simulation are the final snapshot of our previous Enzo simulation of DCBH formation \citep{latif22b} at $z =$ 25 in a 25 Mpc/$h$ periodic box with cosmological initial conditions generated with MUSIC \citep{Hahn11} and cosmological parameters adopted from \textit{Planck} \citep{planck2}.  Five nested grids are centered on the DCBH host halo, which has a mass of $4 \times 10^7$ \Ms, for an effective resolution of $\rm 8192^3$ and a dark matter mass resolution of 3636 \Ms.  During the simulation, we allow up to ten additional refinement levels to resolve gravitational collapse down to scales of $\sim$ 4 pc.   The refinement criteria are based on: (1) a baryonic overdensity threshold of 4; (2) a dark matter mass refinement condition of 0.0625 $\rho_{\rm DM} r^{\ell \alpha}$, where $\rho_{\rm DM}$ is the dark matter density, $r = 2$ is the refinement factor, $\ell$ is the refinement level, and $\alpha = -0.3$ ensures super-Lagrangian refinement; and (3) requiring a minimum of four cells per Jeans length \citep{latif20b, latif22b}.  The 30,000 \Ms\ and 40,000 \Ms\ DCBHs in the original simulation are assumed to merge into a single 70,000 \Ms\ BH because of their proximity.  The BH and its host halo are then evolved down to $z = 10.1$, when the halo has grown to $8.3 \times 10^{10}$ \Ms.

\section{Results} 

% Fig. 3

\begin{figure}
\begin{center}
\begin{tabular}{c}
\includegraphics[scale=0.4]{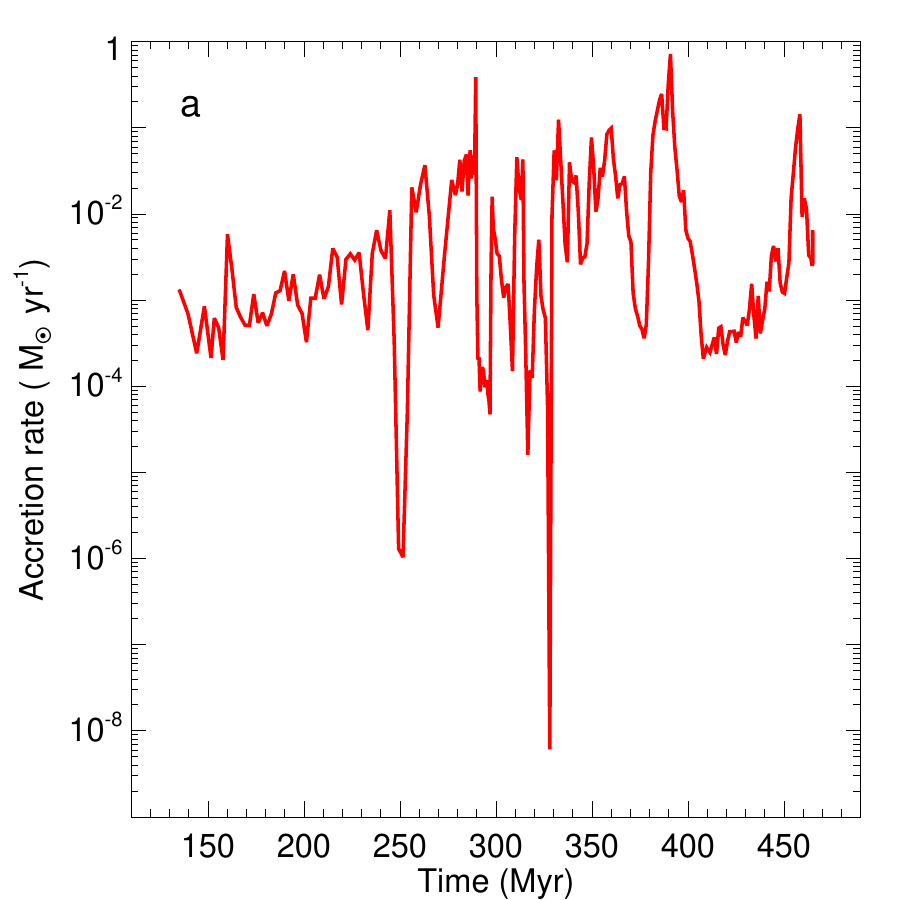}   \\
\includegraphics[scale=0.4]{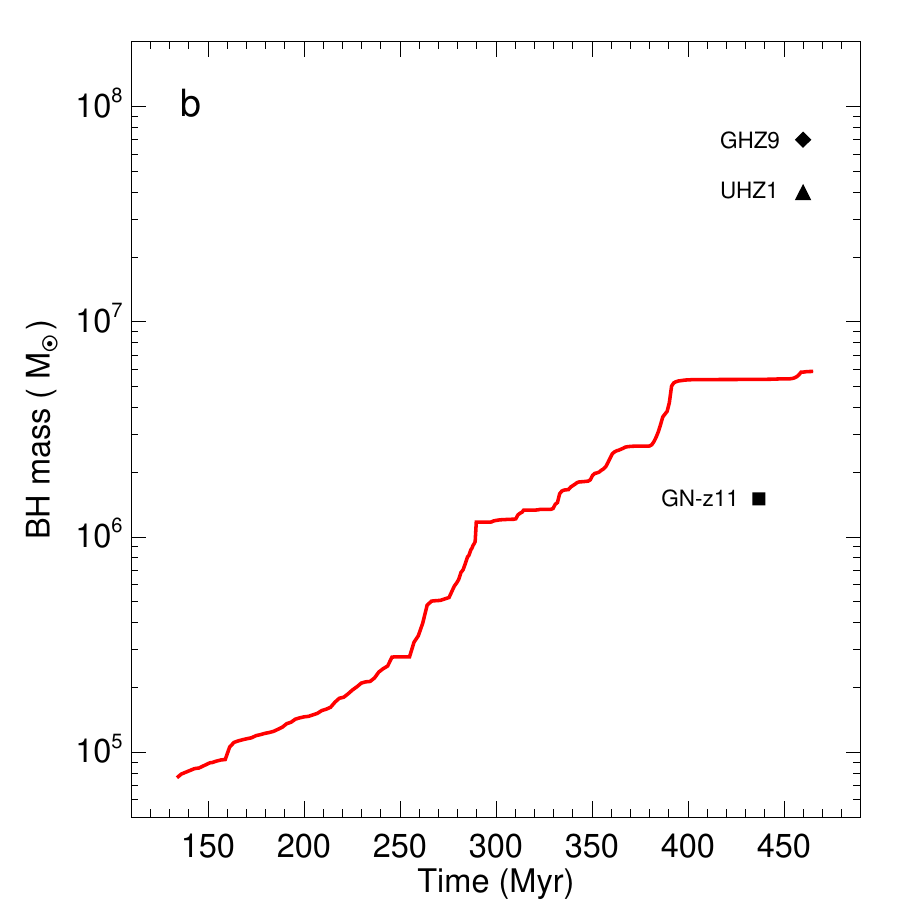}   \\
\includegraphics[scale=0.4]{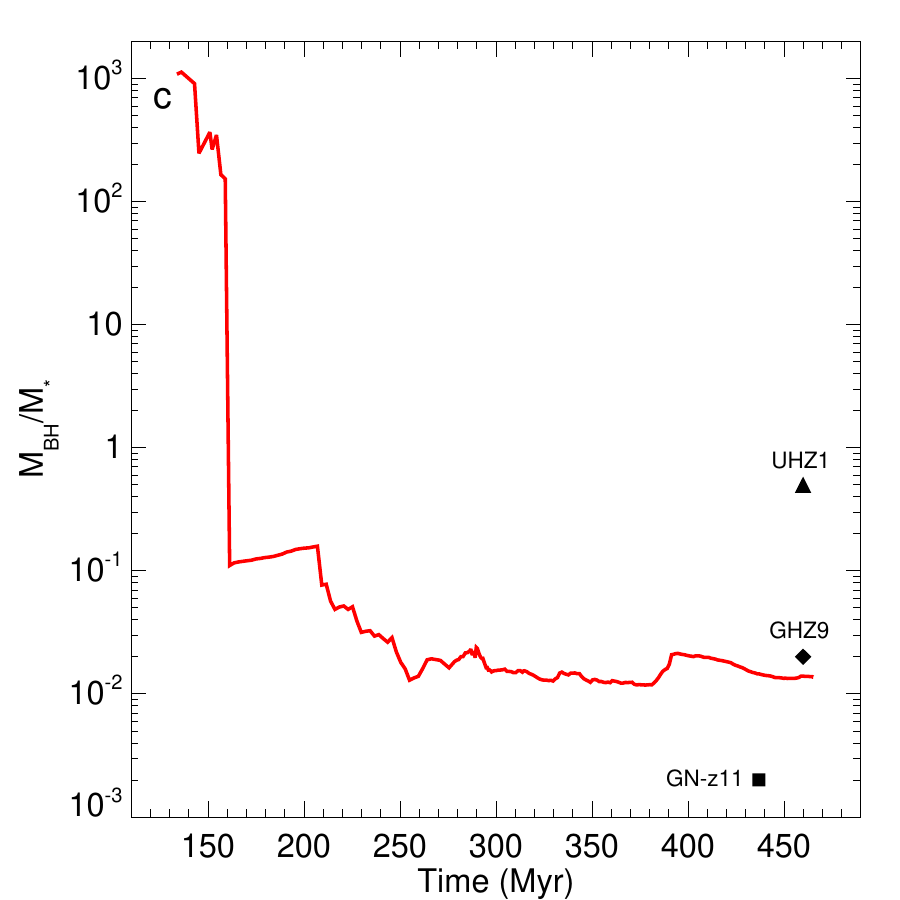} 
\end{tabular}
\end{center}
\caption{BH accretion rate (a), mass (b) and $M_{\rm BH}/M_{*}$ (c) over cosmic time in Myr after the Big Bang.}
\label{fig:BH}
\end{figure}

We show the OBG at birth at $z =$ 25.7 in Figure~\ref{fig:obg}.  X-rays from the BH heat gas in its vicinity, suppressing star formation (SF) in the halo for about 10 Myr after birth, but also elevate $\rm H_2$ mass fractions by more than an order of magnitude, promoting Pop III SF \citep{latif20b}.  After the 10 Myr delay, a 70 \Ms\ Pop III star forms in the dense, cold gas and lives for about 3 Myr before collapsing to a BH.  The formation and death of this star is visible as the first peak in stellar mass at 125 Myr in Figure~\ref{fig:SF}.  No other stars form for the next few Myr. 

Two more Pop III stars, 92 \Ms\ and 249 \Ms, form 2 Myr later near the BH.  One dies in a pair-instability supernova (PI SN) that drives an outflow away from the BH.  SF is then delayed again for $\sim$ 10 Myr, after which it proceeds in a nearby satellite halo that merges with the BH host.  Multiple episodes of Pop~III SF follow, whose SNe enrich the interstellar medium with metals \citep{jet09b,maio11,brit15}.  Once $Z$ exceeds the critical threshold of $\rm 10^{-4}$ \Zs\ as shown in Figure~\ref{fig:SF}, the first Pop II stellar cluster with a mass of $9.5 \times 10^5$ \Ms\ forms near the BH 37 Myr after the onset of SF.  Pop~III SF continues sporadically as the halo is replenished with pristine gas by flows from cosmological filaments.  After the birth of the first Pop~II star cluster, inflow of metal-poor gas intermittently reduces local metallicities below the Pop~II threshold. By 195 Myr after the Big Bang, several Pop~III starbursts of $\sim 10^3$ \Ms\ have again heated the gas but metals cool it within a few Myr because of its high density. As shown in Figure~\ref{fig:SF}, these enrichment events raise $Z$ to $\sim 10^{-3} Z_{\odot}$ over 100 Myr, above the critical value.  Pop~II star formation becomes continuous thereafter, driving the mean metallicity up to $\sim 0.1$ \Zs\ by $z = 10.1$.

The average Pop III star formation rate (SFR) is a few $10^{-4}~\Ms~ \rm yr^{-1}$, peaking at a few $10^{-3}~\Ms~\rm yr^{-1}$ at 280 Myr in Figure~\ref{fig:SF} because of a starburst after a merger with a pristine satellite halo.  Overall, Pop~III SF in the halo is highly bursty because of the stars' short lifetimes.  The mean Pop~II SFR over the last 215 Myr in Figure~\ref{fig:SF} is $\sim$ 2~$\Ms~ \rm yr^{-1}$, whereas over the first 100 Myr it is $\sim$ 0.1~$\Ms~ \rm yr^{-1}$.  The peaks in the Pop II SFR, up to $\sim$ 10~$\Ms~ \rm yr^{-1}$, coincide with halo mergers.  The average total Pop~III stellar mass is a few hundred \Ms, with a maximum of $5.3 \times 10^{3}~ \Ms$ at 310 Myr that coincides with a merger.  The dips in Pop III stellar mass correspond to episodes of PI SNe, which temporarily suppress star formation.  In contrast, the Pop II stellar mass grows continuously over time, reaching $4 \times 10^{8}~ \Ms$ by $z = 10.1$.

As shown in panel (a) of Figure~\ref{fig:BH}, the DCBH accretion rate gradually rises from $\sim 10^{-3}~\Ms~ \rm yr^{-1}$ to $10^{-2}~\Ms~ \rm yr^{-1}$ over the first 125 Myr, or about 50\% of the Eddington rate as shown in panel (b).  X-rays and SNe disrupt but never completely halt accretion because the dense gas cools on timescales of a few Myr, allowing the BH to be refueled, so the growth rate only fluctuates by factors of a few over this time.  They then begin to vary sharply over the next 200 Myr, falling by 2 - 3 orders of magnitude at 250, 290, 315, and 327 Myr because of violent episodes of SF triggered by minor mergers, which drive gas away from the BH.  The average accretion rate of the BH over the simulation is $\sim$ 0.6 Eddington.  The jumps in BH mass in Figure~\ref{fig:BH} correspond to accretion peaks, whereas the plateaus are due to SN-driven outflows that temporarily deplete the local gas supply.  

Tidal torques during the final merger displace the BH from the dense central clump, as in previous studies \citep{Chon21}, dropping its accretion rate to $\sim 10^{-4}~\Ms~ \rm yr^{-1}$ for about 30 Myr before it begins to grow again.  It reaches a final mass of $\sim 6 \times 10^6$ \Ms\ at $z = 10.1$.  The BH-to-stellar mass ratio, $M_{\rm BH}/M_*$, falls to $\sim$ 0.1 when the first Pop~II star cluster forms in the host galaxy.  It then slightly rises over the next 40 Myr as Pop III SNe blow out gas and suppress further Pop II star formation.  Once stars begin to form more rapidly, $M_{\rm BH}/M_*$ gradually falls to $\sim$ 0.01 and remains approximately constant thereafter, apart from small fluctuations associated with peaks in the SFR of the host galaxy shown in Figure~\ref{fig:BH}. $M_{\rm BH}/M_*$ in our simulation is about two orders of magnitude higher than for galaxies today, $M_{\rm BH}/M_{*} \sim 10^{-4}$ \citep{rv15}.  

UHZ1 has a $4 \times 10^7$ \Ms\ BH at $z = 10.1$, stellar mass of $1.4 \times  10^8~ \Ms$, SFR of 1.25 \Ms\ yr$^{-1}$, $M_{\rm BH}/M_{*} = 0.05 - 1$ and $Z \sim 0.2~\Zs$ \citep{Bod23,Gould23}. Similarly, GHZ9 hosts a $\rm 8 \times 10^{7}~\Ms$ BH at z $\sim$ 10, stellar mass of $0.5 - 3.4 \times 10^8~ \Ms$, SFR of $0.5 - 14~ \Ms$ yr$^{-1}$, $M_{\rm BH}/M_{*} \sim 0.02$  and $Z < 0.1~\Zs$ \citep{Atek23,Cast23,Kov24}. GNz11 is an exceptionally luminous galaxy with a $\rm 1.5 \times 10^6~\Ms$ BH at $z =$ 10.6, stellar mass of $\rm 8 \times  10^8 ~ \Ms$, SFR of 20 \Ms yr$^{-1}$, $M_{\rm BH}/M_{*} \sim 0.002$ and $Z \sim 0.2~ \Zs$ \citep{B23,Maio24N}. The BH masses may be uncertain to one or two orders of magnitude because of bolometric corrections and the slopes assumed for their spectral energy distributions, which are taken from objects in the local Universe and may not be applicable to high-$z$ BHs.  Likewise, the stellar masses, SFRs and metallicities are derived from stellar population synthesis codes and are also subject to some uncertainty.  Nevertheless, within these uncertainties the properties of the BH and its host galaxy in our simulation are consistent with those of UHZ1, GHZ9 and GN-z11 as shown in Figures~\ref{fig:SF} and \ref{fig:BH}.  BH masses for GHZ9 and UHZ1 lie somewhat above than those in our simulation but are well within the uncertainties in observed mass.

% Fig. 4

\begin{figure}
\begin{center}
\includegraphics[scale=0.4]{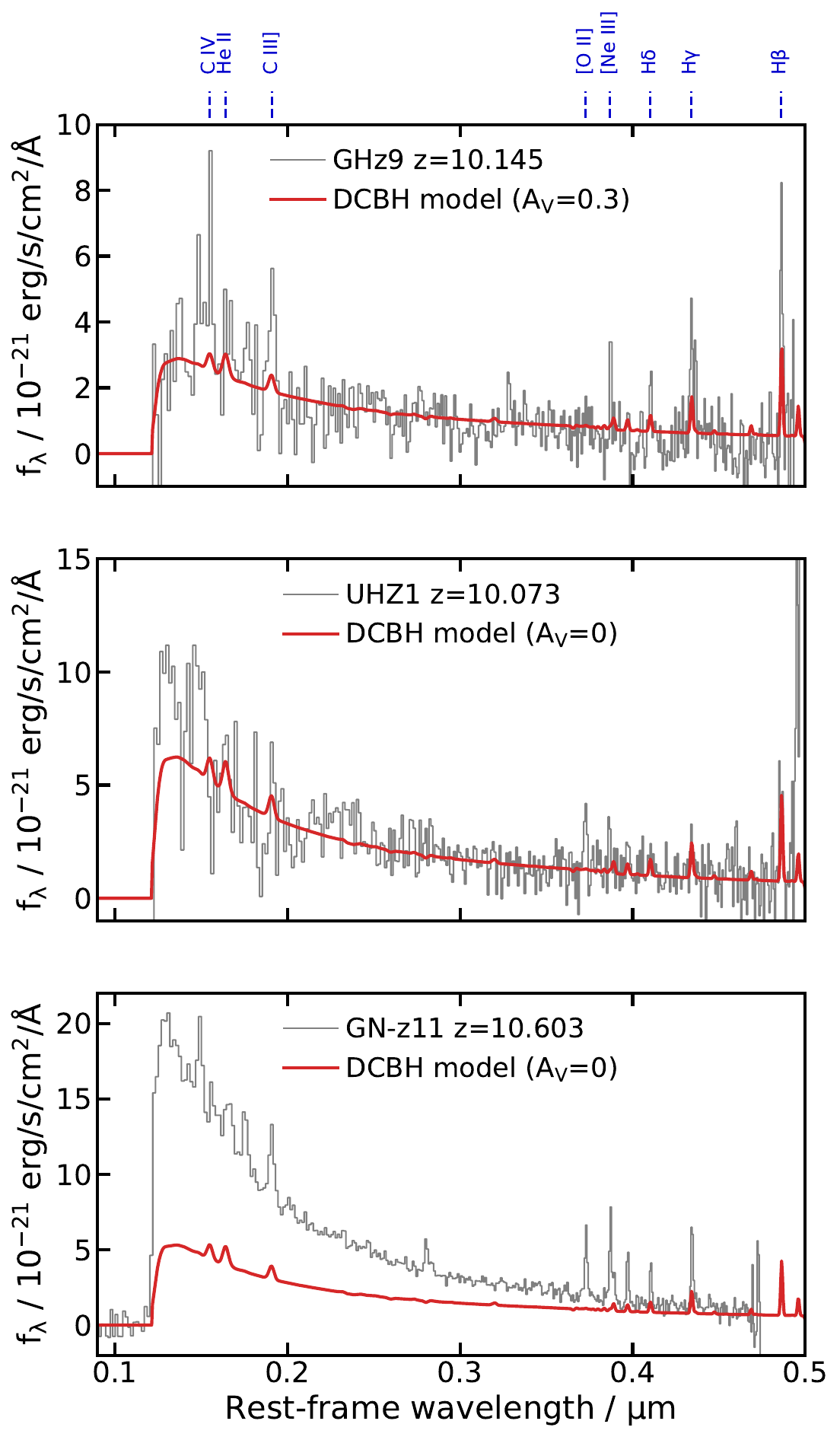} 
\end{center}
\caption{Comparison of Cloudy spectra for the OBG to those for GHZ9, UHZ1 and GN-z11.}
\label{fig:spec}
\end{figure}

We compare a synthetic spectrum for our OBG with the observed spectra for GHZ9, UHZ1 and GN-z11 in Figure~\ref{fig:spec}.  Our spectrum was created with Cloudy v.23.01 \citep{Guna2023}.  We used the ionizing DCBH spectrum from \cite{L25} for the BH, normalized to the appropriate mass, and assumed a spherical geometry with a fixed gas density of $n_{\mathrm{H}} = 10^4 \, \mathrm{cm^{-3}}$ and a metallicity $\mathrm{log}(Z/\mathrm{Z_{\odot}})=-1.7$.  For the stellar population in the host galaxy we used the Binary Population and Stellar Synthesis (BPASS) v.2.3 models \citep{Eld2017, Byrne2022} and constructed a global SED based on the masses, ages and metallicities of the individual star particles, again assuming a spherical geometry but with a fixed gas density of $n_{\mathrm{H}} = 10^3 \, \mathrm{cm^{-3}}$ and a metallicity $\mathrm{log}(Z/\mathrm{Z_{\odot}})=-1.7$.  The stellar and DCBH spectra were summed to produced the final output spectrum.  We find that the DCBH dominates the SED at the UV and optical wavelengths probed by JWST observations, accounting for $\simeq 70 \, \%$ of the total flux in this wavelength range.

For comparison to the GHZ9, UHZ1 and GN-z11 spectra, our synthetic spectrum was converted from luminosity into $f_{\lambda}$ at the redshift of each source and resampled and convolved to match the wavelength sampling and resolution of the observations.  Overall, the comparison is very encouraging.
In the case of GHZ9 and UHZ1, the model matches the shape and normalisation of the continuum extremely well.  To achieve our match with GHZ9, a moderate amount of dust attenuation must be assumed ($A_{V} = 0.3$, assuming a \citealt{Calz2000} attenuation curve).  The emission line strengths are not perfectly recovered but the model does predict line emission where it is generally seen.  An imperfect match is not surprising given the simplifying assumptions of our modelling, such as spherical geometry and standard solar-scaled element abundance ratios.  For GN-z11 the synthetic spectrum is not as good a match:  the observed optical flux is about 1.5 times higher, the observed spectrum has a much steeper SED and is $\simeq 3 \times$ brighter than our model at UV wavelengths. Increasing the normalization of the synthetic stellar SED by a factor 7 (equivalent to assuming 7 times more stellar mass) yields a reasonable match to the observed spectrum. It should be noted that the BH to stellar mass ratio for GN-z11 is an order of magnitude lower than the other two sources, making it less overmassive than UHZ1 and GHZ9 and hence an outlier to some degree. GN-z11 may also be at a later stage of its evolution and could be transitioning into a more typical AGN.   

\section{Conclusion}

While our simulations resolve star formation in the host galaxy they do not capture the accretion disk around the BH, which may, depending on BH age and mass, trap some X-rays and limit their effect on local star formation \citep[see, e.g., Figure 1 of ][]{wet21a}.  This may also be why some OBGs are X-ray quiet while X-ray flux has been detected from others like UHZ1.  However, it is difficult to gauge the net effect trapping has on SFRs in the vicinity of the BH and therefore its growth rate, as X-ray heating suppresses SF but secondary ionizations due to energetic photoelectrons catalyze H2 formation and promote it.  Here, our simulations show that the large $M_{\rm BH}/M_*$ of OBGs is due to initial suppression of SF by X-rays from the BH followed by expulsion of gas and dilution of metals by Pop III SNe, which delays the rapid buildup in mass of Pop II stars for about 100 Myr as shown in Figure~\ref{fig:SF}. 

Although the BH in our model formed from turbulent flows, this sequence of events would be expected for DCBHs forming by other means such as strong Lyman-Werner UV backgrounds \citep{bl03,ln06,wta08,rh09b,agarw12,latif13d,agarw14,pat23}, supersonic baryon motions relative to dark matter in halos \citep{hir17,srg17}, dynamical heating \citep{wise19},or dynamical friction heating \citep{Wu26}, not just synchronised halo pairs.  Given that the numbers of OBGs found so far are consistent with previous estimates of DCBH number densities \citep{agarw14,hab16}, our simulations suggest that OBGs may be a natural phase of evolution in most DCBH hosting galaxies and reinforce the case for massive seeds for the first SMBHs in the Universe (see also \citealt{prole26}).  Massive seeds would also introduce a characteristic "bump" in the SMBH mass function, $dN/dln M_{\rm BH}$ vs. $z$ \citep{jeon25b}, which may have been detected by GLIMPSE/JWST \citep{fei25}. We adopted Eddington-limited spherical Bondi-Hoyle accretion without 
including relative velocity, which may affect the mass accretion onto  the black hole. However, the relative velocity of the black hole is of order 10 km/s, comparable to the sound speed, which can occasionally exceed this value due to energy deposition from the black hole. Therefore, we expect that neglecting the relative velocity does not significantly affect our main conclusions.

\begin{acknowledgments}

We thank the anonymous referee, whose constructive comments improved the quality of this Letter.  MAL thanks the UAEU for funding via UPAR grant No. G00005454. SK acknowledges funding via STFC Small Grant ST/Y001133/1. FC acknowledges support from a UKRI Frontier Research Guarantee Grant (PI Cullen; grant reference EP/X021025/1). For the purpose of open access, the authors have applied a Creative Commons Attribution (CC BY) license to any author accepted manuscript version arising from this submission.

\end{acknowledgments}

\bibliography{ref.bib}
\bibliographystyle{aasjournal}

\end{document}